\documentclass[article,pre,
twocolumn,
nobibnotes,
eqsecnum,
amsmath,amssymb,aps
]{revtex4}

\usepackage{balance}
\usepackage[dvipsnames]{xcolor}
\usepackage{epsfig}
\usepackage{amsfonts}
\usepackage{amsmath}
\usepackage{wasysym}
\usepackage{amssymb}
\usepackage{bm}
\usepackage{float}
\usepackage{enumitem}
\usepackage{rotating}
\usepackage{url}
\usepackage{hyperref}

\usepackage{pdflscape}
\usepackage{graphicx}
\usepackage{ulem}

\usepackage[utf8]{inputenc}
\usepackage{bbm} 				
\usepackage{epstopdf}				
\usepackage{verbatim}    			
\usepackage{sidecap}                
\usepackage{indentfirst}

\newcommand{\be}{\begin{equation}}
\newcommand{\ee}{\end{equation}}
\newcommand{\bea}{\begin{eqnarray}}
\newcommand{\eea}{\end{eqnarray}}

\definecolor{armygreen}{rgb}{0.0, 0.5, 0.0}

\begin{document}

\title{Superstatistical Approach to Turbulent Circulation Fluctuations}

\author{
Henrique S. Lima$^{1}$,
Rodrigo M. Pereira$^{2}$,
Luca Moriconi$^{3}$,
Katepalli R. Sreenivasan$^{4,5}${\footnote{Corresponding author: katepalli.sreenivasan@nyu.edu}},
and
Constantino Tsallis$^{1,6,7,8}$}
\affiliation{\\}
\affiliation{$^{1}$Centro Brasileiro de Pesquisas Físicas, Rua Xavier Sigaud 150, Rio de
Janeiro, RJ, 22290-180, Brazil}
\affiliation{$^{2}$Instituto de F\'{i}sica, Universidade Federal Fluminense, 24210-346 Niter\'{o}i, RJ, Brazil}
\affiliation{$^{3}$Instituto de F\'{i}sica, Universidade Federal do Rio de Janeiro, 21941-909, Rio de Janeiro, RJ, Brazil}
\affiliation{$^{4}$Department of Mechanical and Aerospace Engineering, New York University, New York 11201 USA}
\affiliation{$^{5}$Department of Physics and Courant Institute of Mathematical Sciences,
New York University, New York 10012, USA}
\affiliation{$^{6}$Santa Fe Institute, 1399 Hyde Park Road, Santa Fe, NM, 87501, USA}
\affiliation{$^{7}$Complexity Science Hub Vienna, Metternichgasse 8, Vienna, 1030,
Austria}
\affiliation{$^8$Dipartimento di Fisica e Astronomia Ettore Majorana, University of Catania, Italy}


\begin{abstract}
Recent investigations of turbulent circulation fluctuations have uncovered substantial insights into the statistical organization of flow structures and revealed unexpected geometric features of turbulent intermittency. Of particular interest here is the observation that circulation probability distribution functions admit a superstatistical representation, namely a description based on “ensembles of Boltzmann–Gibbs ensembles.” A fundamental phenomenological ingredient of this approach, which serves as a natural starting point for modeling, relies on the strong correlation between the dissipation field and the spatial distribution of elementary circulation-carrying structures, i.e. small-scale vortices. Within the language of superstatistics, this corresponds to characterizing circulation statistics through an appropriate choice of conditioned (Boltzmann-like) distributions and mixing distributions. We show that the superstatistical class of $q$-exponentials, known to have broad applicability in a wide range of multiscale and non-equilibrium systems, provides an accurate description of the observed circulation statistics in homogeneous and isotropic turbulence. This finding opens avenues for exploring the statistical structure of the turbulent cascade in the context of non-extensive statistical mechanics, rooted in the concept of non-additive entropies.
\end{abstract}


\maketitle

\section{Introduction}

From a broad theoretical perspective, turbulent flows are characterized by multiscale interactions between local shear and vorticity, which underlie the transfer of energy from large to small hydrodynamic scales \cite{Tennekes-Lumley,frisch,Jorg-Sreeni}. A closely related and salient feature of turbulent fluctuations is their intermittent character, manifested by the extended tails of non-Gaussian probability distribution functions of dynamical observables such as velocity increments and velocity-gradient components \cite{batch-town,Jorg-Sreeni,frisch}.


It is noteworthy that a wide class of systems exhibiting similar large fluctuations has been successfully discussed within the framework of non-extensive thermodynamics \cite{tsallis,tsallis2,umarov-tsallis} or, more comprehensively, superstatistics, a theory that describes superpositions of equilibrium ensembles associated with fluctuating intensive parameters \cite{beck, beck-cohen, beckBJP, hanel_etal}.~Even though some results in the turbulence literature exhibit 
correspondences with superstatistical forms \cite{castaing_etal, kholm_etal, apol_etal}, {there remains a conceptual gap because the grounding for superstatistics has remained {\it ad hoc}} \cite{Beck2001CouetteTaylor,beck2,jung_swinney,sosa_etal,gravanis_etal,fried_etal}.
~The present work aims to narrow this conceptual gap within the rich phenomenological setting of circulation statistics \cite{migdal,Iyer_etal,Iyer_etal2,apol_etal}, thereby providing new insight into the scaling properties commonly observed in fluid turbulence.

In this connection, a major modeling challenge, still largely unresolved, is the formulation of the complex phenomenology of turbulence in terms of the emergence and dynamics of specific flow structures. Particular attention has been devoted to homogeneous and isotropic turbulence (HIT) -- our main focus in this work -- where small-scale vortex tubes, with characteristic sizes on the order of the Kolmogorov dissipation length \cite{frisch,afonso_etal,afonso2_etal}, are preferentially generated and amplified by instability mechanisms in the vicinity of thin shear layers.~These layers, in turn, are quasi-two-dimensional regions (viscosity-regularized vortex sheets {or rolled up tubes}) in which turbulent kinetic energy is intensely dissipated \cite{sreeniRMP, ishihara_etal, jorg_etal, elsinga_etal}.


It is thus natural to analyze velocity-gradient fluctuations by separating them into symmetric (dissipation-related) and antisymmetric (rotation-related) components. As a matter of fact, dissipation intermittency has been traditionally discussed in the light of multiplicative cascade models \cite{frisch,O62,K62}, which, when supplemented by Kolmogorov's refined similarity hypothesis, allow one to recover the multifractal scaling behavior of velocity structure functions \cite{FP}. One may wonder, however, in pursuit of a more complete picture of turbulence, about the missing ingredient of rotation, or, more precisely, how the vorticity field is spatially distributed and correlated with the dissipation field. In this regard, it is worth noting that transverse velocity increments, being more directly influenced by rotational structures, display scaling properties that are not strictly identical to their longitudinal counterparts, especially at higher orders, reflecting their enhanced sensitivity to vorticity fluctuations \cite{ChenSreeni}.

A promising direction of study is then suggested by visualizations of the vorticity field obtained from direct numerical simulations (DNS) of turbulence. Having in mind that the turbulent vorticity field is predominantly organized in the form of small vortex tubes \cite{orszag_etal, farge_etal, kaneda_etal,
kaneda_etal2, afonso_etal, afonso2_etal}, a very convenient mathematical probe to investigate their distribution and fluctuations is the velocity circulation $\Gamma$ around an arbitrarily oriented contour $C$,
namely,
\begin{equation}
\Gamma[C] = \oint_C dx_i v_i  = \iint_S \mathbf{\omega} \cdot d\mathbf{A} \ , \  \label{circ}
\end{equation}
where the second relation is the Stokes theorem, $S$ being an oriented surface bounded by
$C$. 

The importance of considering circulation as a key observable in the statistical theory of turbulence was already emphasized by Migdal some three decades ago \cite{migdal}, who proposed a direct connection between circulation statistics and the geometry of minimal surfaces spanned by the integration contours. In the same context, the far-tail structure of circulation probability distribution functions (circulation PDFs) was also discussed. After a relatively long period laced with inconclusive efforts, primarily attributable to limited computational and experimental resources, the study of circulation statistics has flourished in recent years, driven by the extensive DNS investigations of Iyer et al.~\cite{Iyer_etal,Iyer_etal2}, which have revisited Migdal's original conjectures, along with their later, more detailed developments \cite{migdal2}, while also revealing the bifractal nature of circulation fluctuations.~Since then, theoretical, numerical, and experimental efforts have considerably broadened the scope of research, extending from three- and two-dimensional HIT \cite{apol_etal, bounded_measures, mori_etal,mori_pereira,mori_etal2,iyer_mori,mori_pereira2,optimal_surfaces, zhu_etal,muller-krst} to wall-bounded turbulent flows \cite{mug-thor, duan_etal}, and even to quantum turbulence \cite{muller_etal, polanco_etal, muller_etal2, muller-krst2}.

On the phenomenological modeling front, the role of the dissipation field in shaping circulation fluctuations emerges as an essential element of the {\it{vortex gas model}} (VGM) of circulation statistics \cite{apol_etal, bounded_measures, mori_etal,mori_pereira,mori_etal2,iyer_mori,mori_pereira2,optimal_surfaces}.~The core idea of the VGM is that the dissipation field $\epsilon(x)$ is tightly linked to the density field $\xi(x)$ of elementary vortex structures (vortex tubes), which carry characteristic circulations $\tilde {\Gamma}(x)$ around the position $x$. 


The VGM gives explicit modeling definitions for the correlation functions of the dissipation and the elementary circulation fields, so that, in principle, statistical properties of Eq.~\ref{circ} can be obtained and compared to the results of numerical simulations. The dissipation field, specifically, has been described in the VGM along the lines of the Gaussian Multiplicative Chaos (GMC) theory of multifractality \cite{GMC,pereira_etal}, a field-theoretical generalization of the Obukhov-Kolmogorov (OK62) model of intermittency \cite{O62,K62}. 

Although the VGM predictions are consistent with observations \cite{apol_etal,mori_etal,mori_pereira2,optimal_surfaces}, alternative modeling prescriptions remain both worthwhile and necessary.~In particular, the GMC framework characterizes the coarse-grained dissipation field as a lognormal random variable across {\it all} relevant flow length scales. However, it is well established that this assumption breaks down near the dissipation range, where chi-squared and stretched exponential distributions provide, respectively, more appropriate statistical descriptions of small and large fluctuations of the local dissipation field \cite{pk_etal}.

As we shall argue on the basis of superstatistical considerations, this validity of the chi-squared and stretched exponential models is crucial for deriving closed-form analytical expressions for the circulation PDFs, which constitute the central result of our analysis.

This paper is organized as follows. In Sec.~2, we review the technical foundations of the VGM for turbulent circulation, culminating in an integral representation of the circulation PDFs. In Sec.~3, we recall the main ideas of superstatistics \cite{beck, beck-cohen}, thereby establishing a heuristic bridge between the VGM and the alternative approach to circulation statistics to be explored here, referred to as the {\hbox{“$q$-VGM.”}} This formulation relies on the superstatistical derivation of $q$-exponentials \cite{tsallis,tsallis2,umarov-tsallis}.~In Sec.~4, we present systematic (and remarkably accurate) comparisons between the predictions of the {\hbox{$q$-VGM}} and circulation PDFs obtained from DNS databases. Finally, in Sec.~5, we discuss our findings and outline directions for future research.

\section{VGM Essentials}

A concise account of the fundamental technical structure of the VGM is presented below. Let $x$ denote Cartesian coordinates parametrizing a planar region immersed in a three-dimensional HIT flow. Assuming that the circulation may be evaluated via Stokes theorem as the total vorticity flux produced by vortex structures, we rewrite Eq.~\ref{circ} as
\begin{equation}
\Gamma[C] = \int_{\mathcal{D}} dN(x)\,\tilde{\Gamma}(x) \ , \ \label{circ2}
\end{equation}
where ${\mathcal{D}}$ denotes the planar domain enclosed by the contour $C$, and
\vspace{-0.1cm}
\begin{equation}
dN(x) \equiv d^2x\,\frac{\xi(x)}{\eta^2} \ . \ \label{dN}
\end{equation}
The quantity $dN(x)$ represents the expected number of vortex tubes intersecting the infinitesimal area element $d^2x$ inside $\mathcal{D}$, each contributing an elementary circulation $\tilde{\Gamma}(x)$. Accordingly, $\xi(x)$ may be interpreted as the local surface density of vortex-tube intersections, measured in units of $\eta^{-2}$, where $\eta$ denotes the Kolmogorov dissipation scale \cite{frisch}.


The stochastic fields $\xi(x)$ and $\tilde{\Gamma}(x)$ are assumed to be statistically independent. Phenomenologically, the vortex density is observed to scale with the square root of the local energy dissipation rate $\epsilon(x)$ \cite{mori_pereira}. One writes, within the GMC formalism, that
\begin{equation}
\xi(x) \equiv \xi_0 \exp\!\left[\sqrt{\frac{\pi\mu}{2}}\,\phi(x)\right]
\propto \sqrt{\epsilon(x)/\epsilon_0} \ , \label{xi}
\end{equation}
where $\epsilon_0 = \langle \epsilon(x) \rangle$, and $\mu = 0.23 \pm 0.05$
is the intermittency exponent characterizing the power-law decay of the dissipation-field correlator \cite{SreenivasanKailasnath1993, tang_etal}. The scalar field $\phi(x)$ is taken to be Gaussian, with covariance
\begin{equation}
\langle \phi(x)\phi(x') \rangle =
\frac{1}{(2\pi)^2}
\int \frac{d^2k}{k^2}\,e^{ik\cdot(x-x')} \ ,
\end{equation}
and $\xi_0$ is a dimensionless normalization constant. The above integral is understood to be regularized by ultraviolet and infrared cutoffs at $k_\eta = 1/\eta$ and $k_L = 1/L$, respectively, $L$ being the integral scale of the flow.

The elementary circulation field $\tilde{\Gamma}(x)$ is modeled independently as a Gaussian random field \cite{comment1}. 
Its two-point correlation function is defined as
\begin{equation}
\langle \tilde{\Gamma}(x)\tilde{\Gamma}(x') \rangle
=
\tilde{\Gamma}_0^2
\frac{\eta^\alpha}{2\pi \Gamma(\alpha)}
\int d^2k\, k^{\alpha-2}
e^{ik\cdot(x-x') - k\eta} \ , \ \label{gamma-gamma1}
\end{equation}
which ensures $\langle \tilde{\Gamma}^2 \rangle = \tilde{\Gamma}_0^2$. Furthermore, in Eq.~\ref{gamma-gamma1}, $\Gamma(\alpha)$ denotes the Euler Gamma function. For inertial range separations, {\hbox{$\eta \ll |x-x'| \ll L$}}, this correlator behaves as
\begin{equation}
\langle \tilde{\Gamma}(x) \tilde{\Gamma}(x') \rangle
\sim |x-x'|^{-\alpha} \ , \label{gamma-gamma}
\end{equation}
where
\be
\alpha = 2 - \frac{\mu}{4} - \zeta_2 \ , \ 
\ee 
and $\zeta_2 \approx 2/3$ is the scaling exponent of second order velocity structure functions \cite{frisch}.

An interesting consequence \cite{apol_etal, mori_pereira} of the VGM phenomenological postulates Eqs.~\ref{xi}-\ref{gamma-gamma} is that fluctuations of $\Gamma[C]$ can be effectively expressed as
\be
\Gamma[C] = X[C] \cdot Y[C]  \ , \label{XY}
\ee
where
\be
X[C] \propto 
\int_{\mathcal{D}} d^2x \, \tilde \Gamma(x) \label{X}
\ee
and
\be
Y[C] \propto  \int_{\mathcal{D}} d^2x \, \xi(x) \label{Y} 
\ee
are, respectively, Gaussian and lognormal random variables associated to the variances $\sigma^2_X$ and $\sigma^2_{\ln(Y)}$. It is not difficult to show, from Eqs.~\ref{XY}-\ref{Y}, that circulation PDFs can be recast in the integral form
\be
p(\Gamma) =  \int_0^\infty d \zeta \, f(\zeta) \exp \left ( - \zeta \Gamma^2 \right ) \ , \label{cPDF}
\ee
where
\be
f(\zeta) = \frac{1}{\sqrt{8 \pi}\sigma_{\ln(Y)}} \frac{1}{\sqrt{\pi \zeta}} \exp \left \{  \frac{\left [\ln(\zeta) - \ln(\zeta_0) \right ]^2}{8 \sigma^2_{\ln(Y)}}  \right \} \label{fzeta}
\ee
with
\be
\ln(\zeta_0) \equiv -2 \langle \ln(Y) \rangle - 4 \ln \left (\sigma_X \right ) \ . \
\ee
It is clear that $f(\zeta) \sqrt{\pi/\zeta}$ is simply the lognormal distribution of $\zeta$. The structure of the integral expression, Eq.~\ref{cPDF}, for the circulation PDF is similar to what one finds from the general superstatistical formulation of random observables, which has been widely applied to complex systems of completely distinct phenomenologies \cite{beckBJP}.
This apparently casual observation becomes the guiding idea for a physically meaningful variant of the VGM, as we discuss in the next section.


\section{Superstatistics and the $q$-VGM}

The integral representation of the circulation PDF, Eq.~\ref{cPDF}, has the form of a superposition of Boltzmann weights. This structural feature suggests a natural generalization of the VGM in which the {\it{mixing distribution}} $f(\zeta)$ and the {\it{circulation energy}} $E(\Gamma) \equiv \Gamma^2$ are modified in a physically motivated way, leading to the very compact formulation of $q$-exponential circulation statistics.

Superstatistics \cite{beck,beck-cohen,beckBJP} describes non-equilibrium systems characterized by a separation of scales. On short spatiotemporal scales, a fluctuating variable $x$, with associated energy $E(x)$, is locally described by a Boltzmann--Gibbs distribution
\begin{equation}
p(x | \zeta) \propto e^{-\zeta E(x)} \ , \
\end{equation}
where $\zeta$ is an intensive parameter (inverse temperature, or, more generally, inverse variance). On much longer scales, however, $\zeta$ itself fluctuates according to a probability density $f(\zeta)$. The marginal (non-normalized) distribution of $x$ is then obtained by averaging over $\zeta$, viz.,
\begin{equation}
p(x) \propto \int_0^{\infty} d\zeta \, f(\zeta)\,
 e^{-\zeta E(x)} \ . \
\label{eq:superstat_general}
\end{equation}
Equation \ref{eq:superstat_general}, which can be regarded as a Laplace transform of the distribution $f(\zeta)$, has the same structure as Eq.~\ref{cPDF} for the circulation PDF, with $\Gamma$ playing the role of $x$ and $\zeta$ interpreted as an effective inverse variance. The underlying hypothesis of time scale separation between strain and vorticity (and hence circulation) is actually supported by the phenomenological relevance of vortex stretching dynamics in the processes of eddy (vortex tube) production and evolution \cite{Tennekes-Lumley}.

In the original VGM construction, $f(\zeta)$ is related to a lognormal distribution as a consequence of the GMC description of fluctuating dissipation according to the 1962 description by Obukhov and Kolmogorov. However, detailed small-scale analyses reveal systematic deviations from pure lognormal behavior.~In particular, as one approaches the dissipation range, the statistics of the local dissipation field display features more consistent with chi-squared–type cores and stretched-exponential tails \cite{pk_etal}. From the superstatistical standpoint, this suggests reconsidering the specific choices of the mixing distribution and the circulation energy, while keeping intact the general integral structure of the circulation PDFs.

A particularly interesting choice for the mixing function is the Gamma distribution,
\begin{equation}
f(\zeta) =
\frac{1}{\Gamma(k)}
\left(\frac{k}{\zeta_0}\right)^k
\zeta^{k-1}
\exp\!\left(-\frac{k\zeta}{\zeta_0}\right) \ , \
\qquad k>0 \ , \
\label{eq:gamma_dist}
\end{equation}
which has mean of $\zeta$ = $\langle \zeta \rangle = \zeta_0$ and relative variance
\begin{equation}
\frac{\langle(\zeta-\zeta_0)^2 \rangle}{\langle \zeta \rangle^2}
= \frac{1}{k} \ . \ \label{var}
\end{equation}
We recall that the Gamma family naturally arises from sums of squared Gaussian variables 
\cite{feller} which, as commented above, are relevant in discussions of dissipation intermittency.
Substituting Eq.~\ref{eq:gamma_dist} into Eq.~\ref{eq:superstat_general} and considering, for simplicity, energies $E(x) \equiv |x|^h$, with $h>0$, one finds
\begin{align}
p(x)
&\propto
\int_0^{\infty} d\zeta \,
\zeta^{k-1}
\exp\!\left[-\zeta\left(|x|^h + \frac{k}{\zeta_0}\right)\right] \nonumber \\
&\propto
\left[1 + \frac{\zeta_0}{k} |x|^h  \right]^{-k} \ . \
\end{align}
Introducing now
\begin{equation}
q = 1 + \frac{1}{k} \ , \
\qquad
\beta = \frac{\zeta_0}{k} \ , \ \label{q-beta}
\end{equation}
the distribution may be written as
\begin{equation}
p(x) \propto
\left[1 + (q-1) \beta |x|^h \right]^{-\frac{1}{q-1}}
\equiv e_q^{-\beta |x|^h} \ , \
\label{eq:qexp_superstat}
\end{equation}
where the $q$-exponential function is defined as
\begin{equation}
e_q^{-y} =
\left[1 + (q-1)y\right]^{-\frac{1}{q-1}} \ , \
\qquad (q>1) \ . \
\end{equation}
Hence, Gamma-distributed inverse-temperature fluctuations lead naturally to the stretched-tailed $q$-exponential distributions.
\begin{figure*}[t]
    \centering   
    \includegraphics[width=8cm]{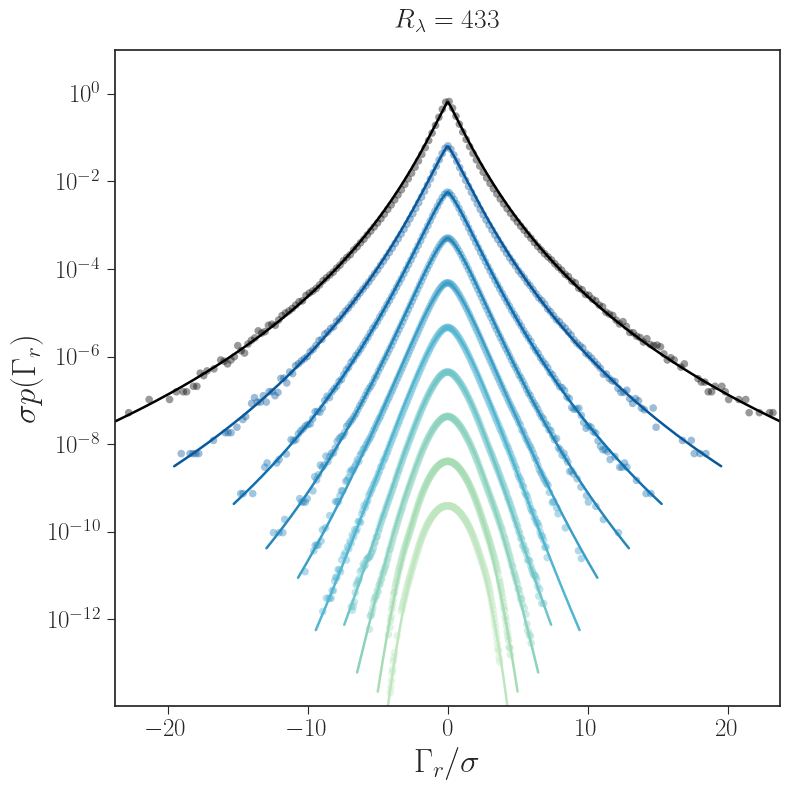}
    \includegraphics[width=8cm]{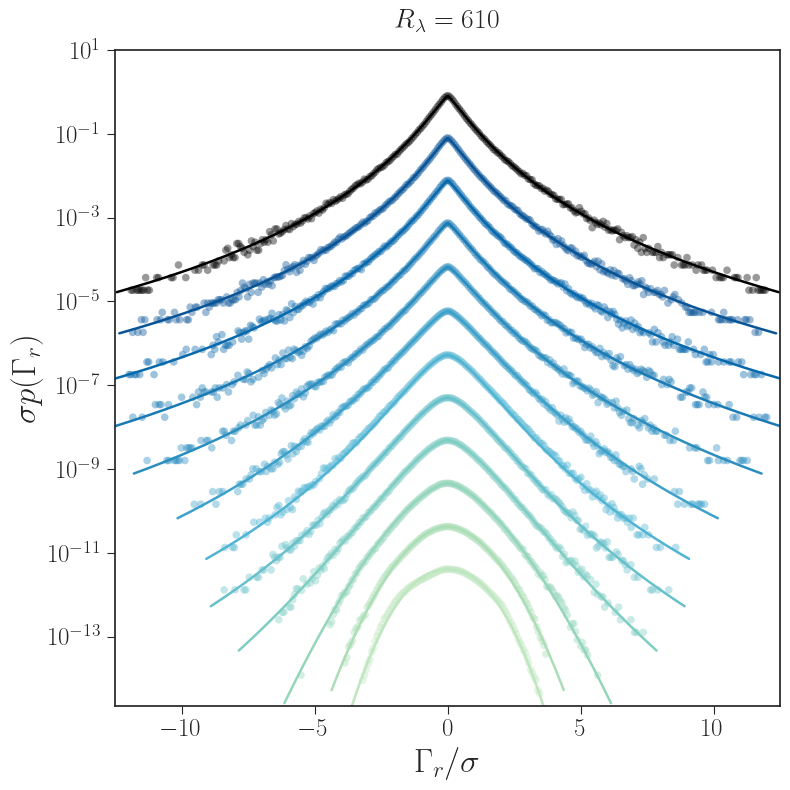}
    \includegraphics[width=8cm]{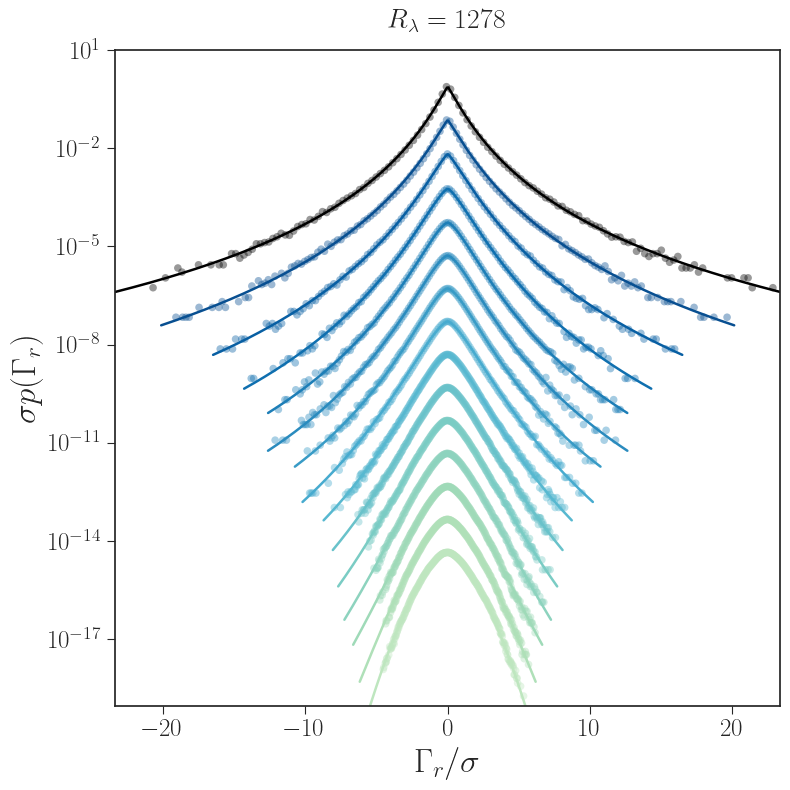}
    \includegraphics[width=8cm]{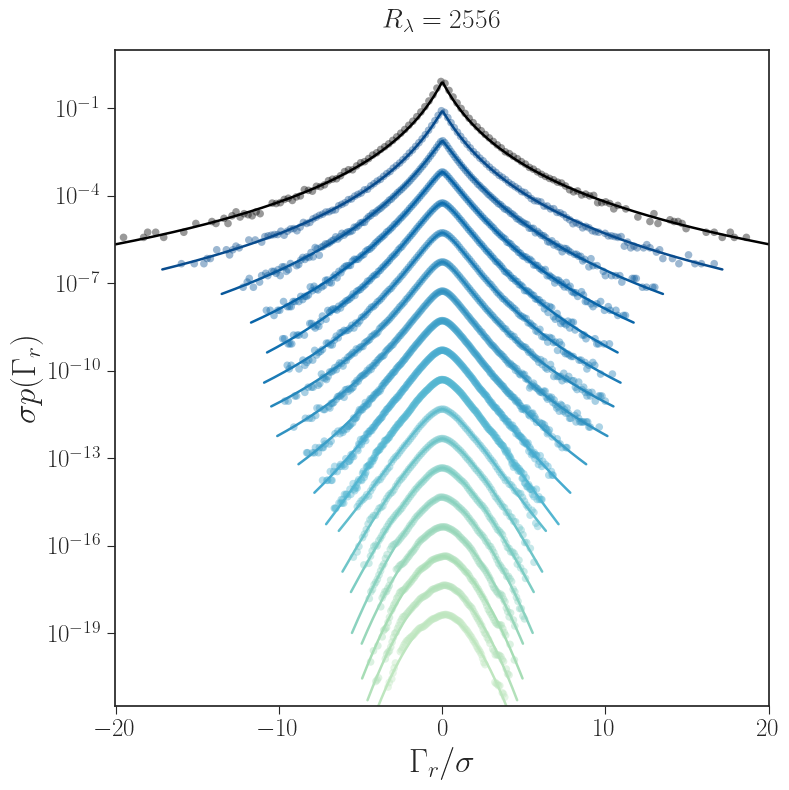}
    \label{}
    \caption{Standardized circulation PDFs evaluated over square contours of sides $r$ that run from deep in the dissipative range ($r \approx \eta$) up to the top of the inertial range. The scale $r$ increases from top to bottom across the PDFs. Symbols denote DNS data, while solid lines correspond to $q$-exponential fits. For visual clarity, the curves are vertically offset.}
\end{figure*}

The same functional form arises independently within the deeper foundational setup of nonextensive statistical mechanics \cite{tsallis, tsallis2, umarov-tsallis}. Consider the nonadditive entropy
\begin{equation}
S_q =
\frac{1 - \int dx \,[p(x)]^q}{q-1} \ , \ 
\end{equation}
which reduces to the Boltzmann--Gibbs entropy in the limit $q \to 1$. Maximizing $S_q$ under the constraints
\begin{equation}
\int dx\, p(x) = 1 \ , \
\qquad
\int dx\, p(x)\, |x|^{h} = \mathrm{const} \ , \
\end{equation}
we are led to a distribution which has the exact functional form
as Eq. (\ref{eq:qexp_superstat}). 
Therefore, the stretched-tailed distributions obtained from gamma superstatistics coincide with those derived from the variational principle of nonextensive thermodynamics. In the present context, we adopt the superstatistical interpretation from a heuristic point of view, while devoting our attention to the suggestive equivalence with the nonadditive entropy-maximization route.

\begin{figure*}[t]
    \centering
    \includegraphics[width=0.90\linewidth]{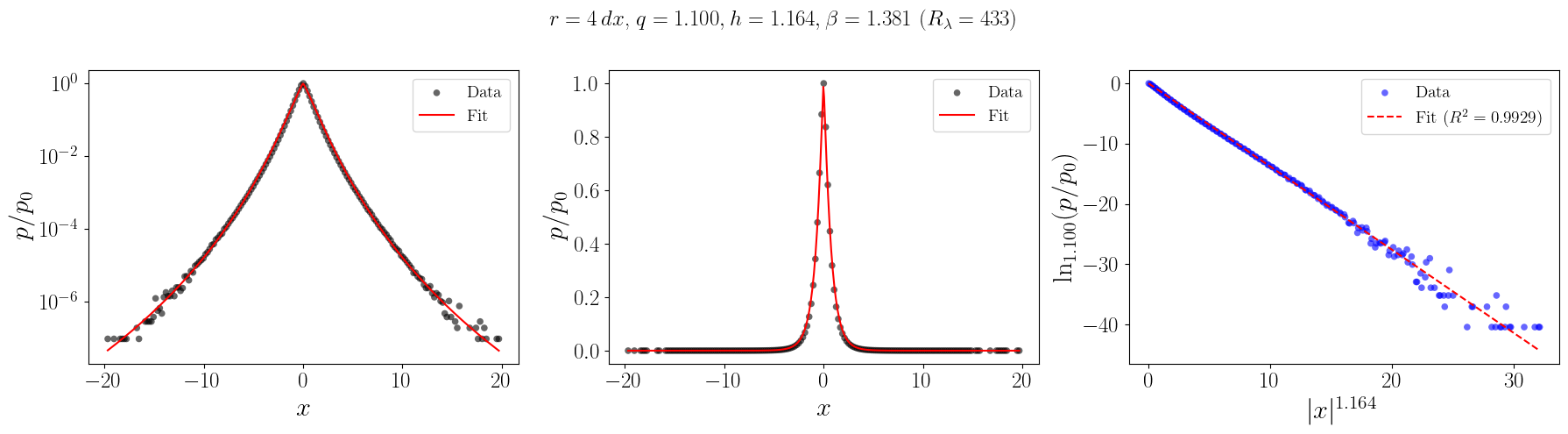}
    \includegraphics[width=0.90\linewidth]{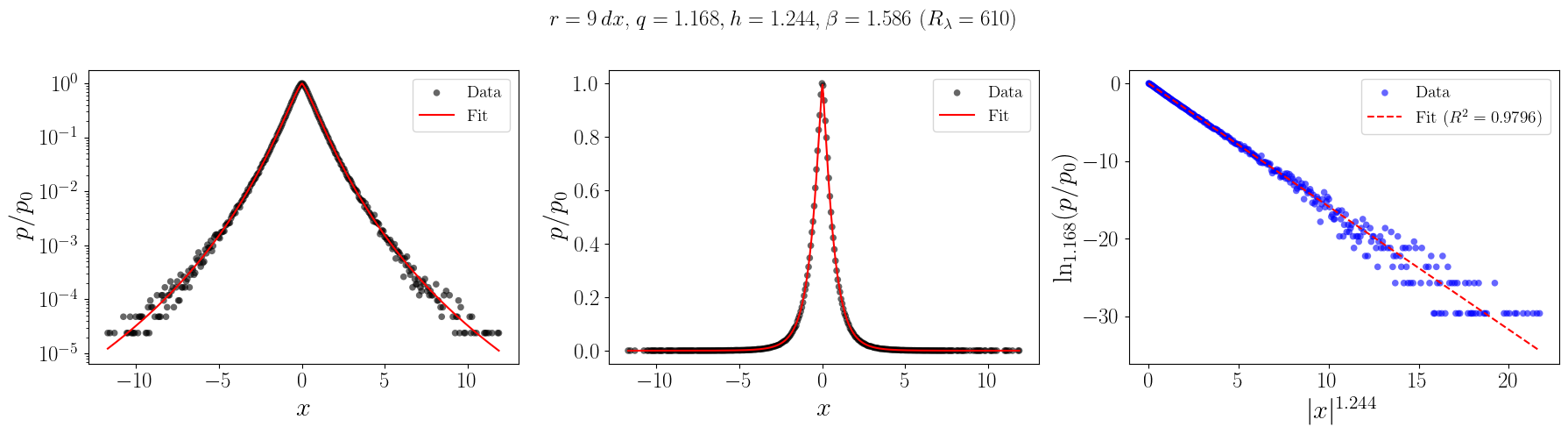}
    \includegraphics[width=0.90\linewidth]{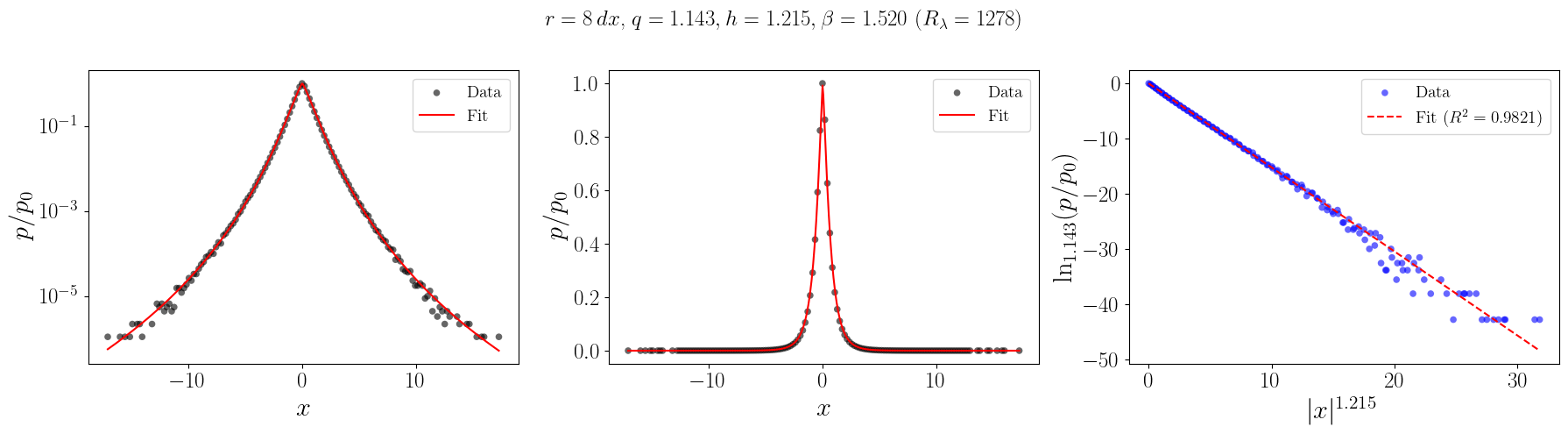}
    \includegraphics[width=0.90\linewidth]{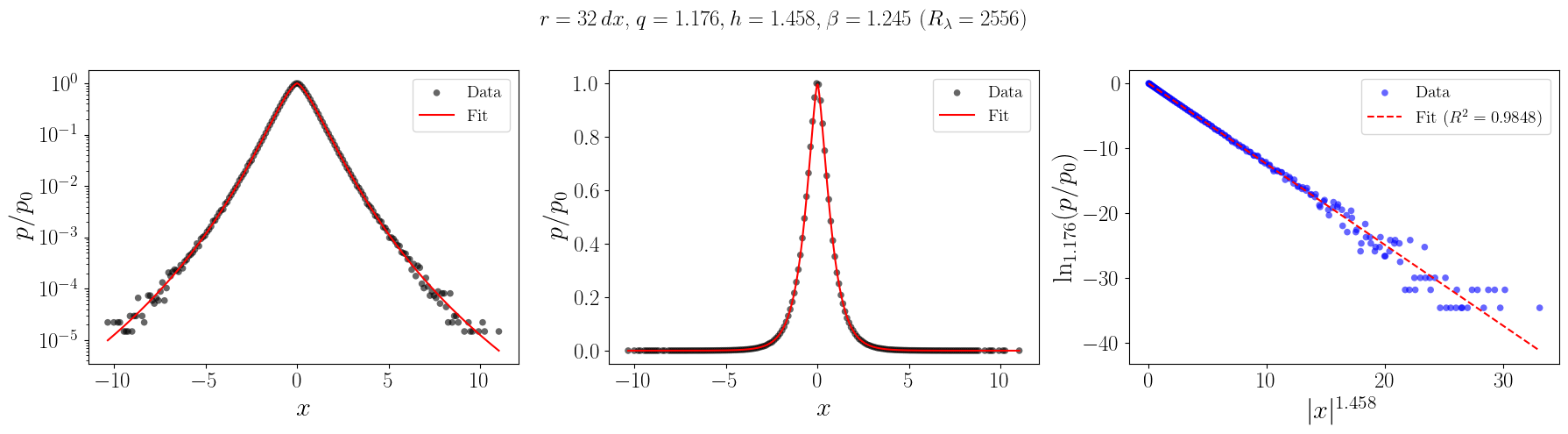}
    \caption{A sample of standardized circulation PDFs, for different Reynolds numbers. A high-quality linear regression ($R^2 \approx 1$) is generally achieved for $\ln_q(p(x)/p(0))$ versus
    $|x|^h$, where $x = \Gamma_r / \sigma$ ($r$ is the side of the square circulation contour, an integer multiple of the grid resolution $dx$, which is typically within the range $0.5 \leq dx/\eta \leq 1$).
}
    \label{fig2}
\end{figure*}

We now introduce the $q$-VGM. Starting from the VGM representation,
Eq.~\ref{cPDF}, we replace the VGM mixing distribution Eq.~\ref{fzeta} and the quadratic circulation energy by, respectively, the gamma distribution, Eq.~\ref{eq:gamma_dist}, and {\hbox{$E(\Gamma) = |\Gamma|^h$}}. The resulting circulation PDF becomes
\begin{equation}
p_q(\Gamma)
=
p(0)
\left[
1 + (q-1)\beta | \Gamma|^h
\right]^{-\frac{1}{q-1}} \ . \ 
\label{eq:qvgm_pdf}
\end{equation}
In this formulation, we have, from Eqs.~\ref{var} and \ref{q-beta},
\begin{equation}
q  = 1 + \frac{\langle(\zeta-\zeta_0)^2 \rangle}{\langle \zeta \rangle^2}  \ , \
\end{equation}
so that $q$ directly quantifies the strength of inverse-variance fluctuations. Note that a stretched exponential (or even Gaussian) form of the circulation PDF may be obtained in the limit $q \to 1$, corresponding to negligible $\zeta$ fluctuations.

The $q$-VGM thus retains the physical core of the vortex gas picture, namely, the coupling between vortex density and dissipation fluctuations, while adopting a superstatistical closure that naturally yields $q$-exponential circulation distributions.~In what follows, we test this formulation against DNS data for HIT.

\begin{figure}[t]
\begin{center}
\includegraphics[width=8cm]{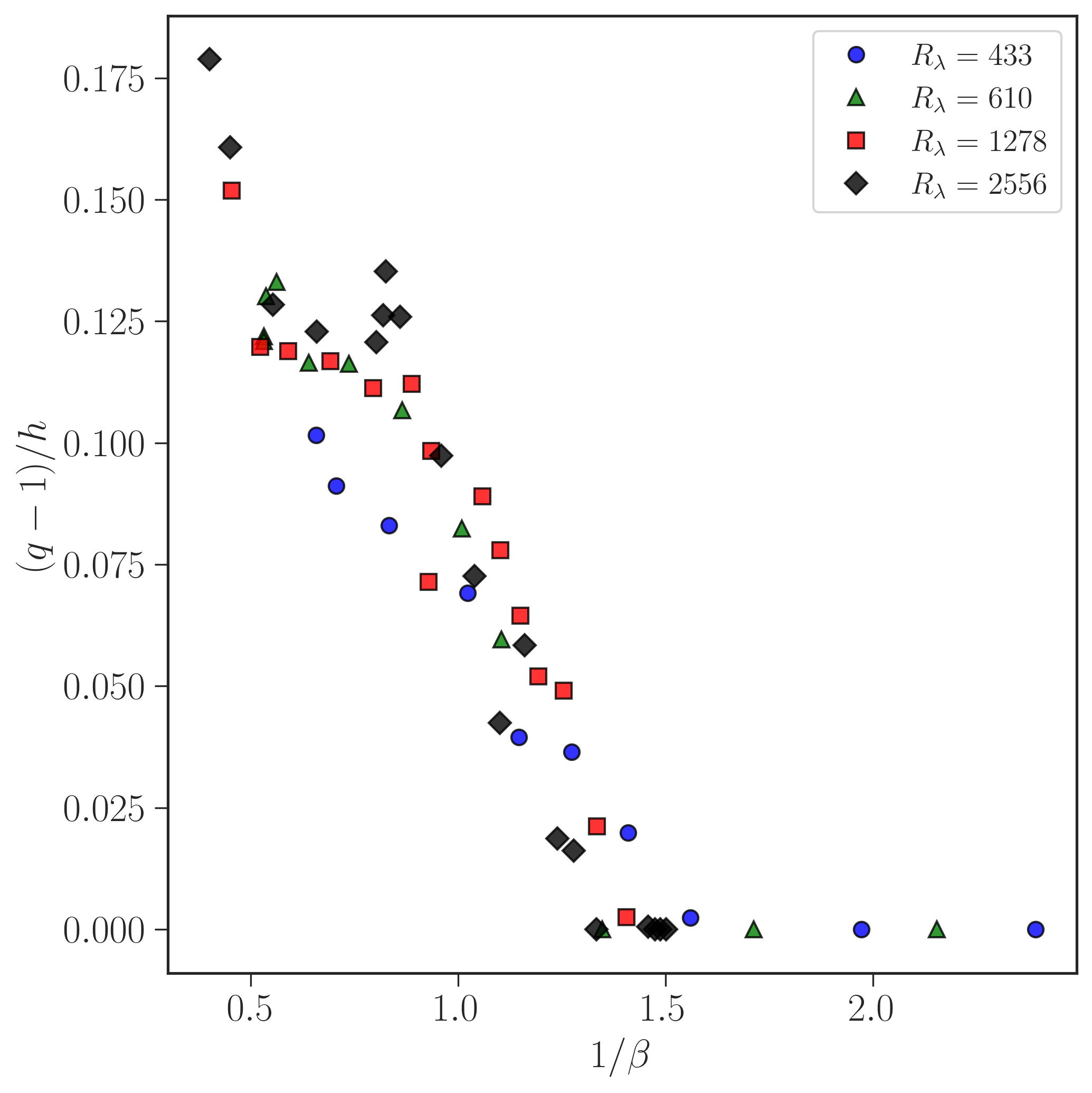}
\caption{\small{Plots of $(q-1)/h$ given as functions of $1/\beta$. The values of $q$, $h$, and $\beta$ are the optimal parameters for the fits shown in Fig.~1. The data collapse, across Reynolds numbers, and the monotonic evolution of $(q-1)/h$ as the sizes of the circulation contours grow (reflected in the growth of $1/\beta$) indicate the existence of critical scaling for the fluctuations of the velocity circulation.}}
\label{figC}
\end{center}
\end{figure}
\section{Circulation Across Scales and Reynolds Numbers}

Our post-processing analysis is based on direct numerical simulation data obtained from the publicly accessible Johns Hopkins University Turbulence Database \cite{JHTDBsite, JHTD,JHTD2,JHTD3,JHTD4,JHTD5}. Four distinct datasets are considered, with Taylor-scale Reynolds numbers $R_\lambda = 433, 610, 1278$ and $2556$. For the evaluation of probability distribution functions, square circulation contours of varying sizes are examined, comprehensively spanning the corresponding inertial-range scales.

Using (\ref{eq:qvgm_pdf}) for standardized circulation PDFs, the parameters $q$, $h$, and $\beta$ are obtained from best fits through a global optimization strategy using the Tree-structured Parzen Estimator (TPE) algorithm within the Optuna framework~\cite{bergstra2011}.~For each candidate pair $(q,h)$, the parameter $\beta$ is estimated using an Iteratively Reweighted Least Squares (IRLS) approach~\cite{beaton1974}.~This estimation technique employs a  weighting function to minimize the influence of statistical noise in the distribution tails, ensuring that the fit is primarily driven by the core and intermediate regions of the datasets. The final parameters $(q,h,\beta)$ are then obtained by minimizing a composite objective function that balances  $R^2$ and Root Mean Square Error criteria (RMSE). 

The results shown in Figs.~1 and 2 demonstrate the excellent accuracy of the $q$-VGM. In Fig.~1, the circulation PDFs are well described to a good approximation by $q$-exponential distributions for all the Reynolds numbers investigated throughout the inertial range of scales, with several decades of the distributions faithfully reproduced. The quality of these fits is further quantified in {the last column of} Fig.~2, where the coefficients of determination, {\hbox{$R^2 \simeq 0.976 \pm 0.026$}}, confirm the predicted linear dependence of the $q$-logarithm defined from Eq.~(\ref{eq:qvgm_pdf}) as
\be
\ln_q \left ( \frac{p(x)}{p(0)} \right ) \equiv \frac{((p(x)/p(0))^{1-q}-1}{1-q} = - \beta |x|^h \ , \ 
\ee
with respect to $|x|^h$, where we take $x \equiv \Gamma/\sqrt{\langle \Gamma^2 \rangle}$.

As the probed scales \(r\) of the circulation contours increase, we observe that \(h \to h^\ast \approx 1.6\) rapidly for \(r/\eta \gtrsim 200\), while \(\beta\) and \(q\) decrease monotonically, with \(q \to q^\ast \approx 1\).~The data further suggest that \(\beta\) tends toward an inertial-range fixed point in the range \(0.5 < \beta^\ast < 1\) at larger scales.

Despite the use of three independent parameters in the fitting procedure, they are in fact nontrivially interrelated, as evidenced in Fig.~3 by the collapse of the curves of $(q-1)/h$ versus $1/\beta$ across the range of Reynolds numbers investigated. Note, from (\ref{eq:qvgm_pdf}), that the ratio $h/(q-1)$ is the exponent for the power law decay of the far tails of the circulation PDFs. The collapse indicates that the statistical parameters of the circulation distributions lie on a one-dimensional manifold in parameter space: once the contour size is specified, the distribution is effectively determined by a single parameter, which, for instance, can be taken to be $\beta$. This reduction to an effective one-parameter description parallels the behavior of critical thermodynamic systems under renormalization group (RG) flow \cite{Amit1984}.

In our turbulence setting, a small-scale cutoff is prescribed by
the size of the circulation contour.~The data collapse of Fig.~3 suggests that the intermittency mechanism governing circulation statistics simultaneously determines both the mean level and the fluctuations of the effective parameter $\beta$, as the small-scale cutoff is increased by a coarse-graining procedure. This behavior points to a turbulent state characterized by long-range correlations and approximate scale invariance over the inertial range, bearing, additionally, a close analogy with systems that display self-organized criticality \cite{Bak1996,Greco2020,Oliveira2024}.


\section{Discussion}

Relying on vortex-gas modeling ideas, we have investigated the statistics of turbulent circulation with the perspective of superstatistics.~It turns out that the PDFs of circulation for contours of different sizes and Reynolds numbers can be accurately described by {\hbox{$q$-exponential}} distributions, which capture the strongly non-Gaussian character of the fluctuations and provide a compact parametrization of intermittency.~Within the superstatistical interpretation, the parameter $\beta$ in Eq.~\ref{eq:qvgm_pdf} can be viewed as an effective inverse fluctuation intensity associated with local quasi-equilibrium states, while the non-extensive entropic index $q$ quantifies the strength of fluctuations of this intensive parameter.~In this sense, $q$ provides a quantitative measure of intermittency in circulation statistics.

Very notably, the monotonic relation observed between $(q-1)/h$ and $1/\beta$ exhibits a collapse for the data associated with different Reynolds numbers when the control parameter is the size $r$ of the circulation contours. 


A natural way to address this observation is to regard $(q,h,\beta)$ as running parameters under a scale transformation $r \to \lambda r$, with the dimensionless control variable $\ell = r/\eta$.~The empirical collapse of Fig.~3 suggests that they lie on a Reynolds-number-independent invariant manifold defined by an “equation of state” $f(q,h,\beta,\ell)=0$, which should be preserved under the RG flow. 
The observed data collapse implies that the flow is effectively constrained to a one-dimensional trajectory, consistent with a reduced description such as $(q-1)/h = \Phi(1/\beta)$, so that the RG dynamics drives the system along this curve toward an inertial-range fixed point $(q^*,h^*,\beta^*)$ as $\ell$ grows. The monotonic behavior of $(q-1)/h$ with $1/\beta$ then reflects the irreversibility of this flow, suggesting an underlying scaling symmetry that organizes the statistics of circulation in direct analogy with universality classes in critical systems.

From a related field-theoretic standpoint, we note that it would be worth revisiting the GMC model of energy dissipation, which may be interpreted as an inertial-range (coarse-grained) fixed point of a more general effective field theory governing the turbulent cascade at shorter scales.

Furthermore, our results point to the need for a closer integration between non-extensive thermodynamics \cite{tsallis,tsallis2,umarov-tsallis} and non-equilibrium statistical mechanics in the statistical description of turbulent fluctuations \cite{ciliberto_etal,Yao_Zaki_Meneveau_2023}. Progress along this direction will likely require a deeper characterization of the dynamical structures underlying intermittency, including the statistics of dissipation layers and elementary vortices.

\acknowledgments{LM, HSL, and CT acknowledge partial financial support from the Conselho Nacional de Desenvolvimento Científico e Tecnológico (CNPq), while LM and CT also acknowledge support from the Fundação de Amparo à Pesquisa do Estado do Rio de Janeiro (FAPERJ). KRS thanks New York University for its support of his work.}




\begin{thebibliography}{99}
\bibitem{comment1} The circulation carried by the elementary vortices is more accurately modeled as a composite field, which behaves as a Gaussian random field only at length scales larger than the Taylor microscale. In practice, the statistical properties of the circulation are still described by Eqs.~\ref{circ2} and \ref{dN}, since  short-distance behavior of the vortex density field $\xi(x)$ is also corrected for compensation, preserving the integrand of Eq.~\ref{circ2}. See Ref.~\cite{mori_pereira2} for details.
\end{thebibliography}

\end{document}